\documentclass{PoS}

\usepackage{sidecap}
\usepackage{gensymb}

\newfont{\rmxs}{cmr10 at 6.3pt}
\newfont{\itxs}{cmti10 at 6.3pt}

\title{A parameterized catalog of radio galaxies as ultra-high energy cosmic ray sources}

\ShortTitle{Catalog of Radio Galaxies as UHECR Sources}

\author{\speaker{J\"org P. Rachen}\\
        Astrophysical Institute, Vrije Universiteit Brussel, Pleinlaan 2, 1050 Elsene, Belgium\\
        E-mail: \email{Jorg.Paul.Rachen@vub.be}}

\author{Bj\"orn Eichmann\\
        Ruhr Astroparticle and Plasma Physics Center (RAPP Center), Ruhr-Universit\"at Bochum, Institut f\"ur Theoretische Physik IV, 44780 Bochum, Germany\\
        E-mail: \email{eiche@tp4.rub.de}}

\abstract{Many attempts have been made to provide catalogs of potential sources of ultra-high energy cosmic ray (UHECR) particles based on various astronomical tracers, such as observed radio or gamma-ray emission. A closer look reveals, however, that they all suffer from significant bias and selection effects. We present here a demo-version of a catalog for one often-discussed UHECR source class, radio galaxies (or radio-loud AGN), which is based on a complete theoretical description of jet-energetics, particle acceleration physics, relativistic beaming effects and nuclear composition, parametrized by a comprehensible set of adjustable physical quantities. In addition to the bright radio galaxies Centaurus A, Virgo A, Fornax A and Cygnus A discussed in previous work, we find several sources with blazar-like properties that can contribute on a similar level if relativistic beaming effects are considered. We present a simple method to test the basic properties of the catalog for any choice of parameters (without the need to run expensive simulations), and find that in the canonical case the anisotropy signals expected from radio galaxies promise to be in good agreement with current observational findings. In particular, radio galaxies can reproduce almost exactly the direction of the dipole above 8\,EeV detected by the Pierre Auger Observatory if we assume that extragalactic magnetic fields are strong (${\gtrsim}\,1$nG) only in structures, but very weak in voids. We plan to provide a completed and improved version of this catalog in electronic form, to be used in more detailed UHECR propagation simulations. For immediate applications, we suggest a complete set of 16 strong UHECR sources which can contribute to UHECR anisotropy on the level which can be currently probed by experiment, and note that 6 of them have not been considered in any previous studies.    
}

\FullConference{36th International Cosmic Ray Conference -ICRC2019-\\
		July 24th - August 1st, 2019\\
		Madison, WI, U.S.A.}

\begin{document}

\section{Introduction}

\noindent Ever since evidence for the existence of ultra-high energy cosmic ray (UHECR) particles up to about 100\,EeV has been achieved, radio galaxies (RGs) belonged to the prime candidates for astrophysical sources able to explain them \cite{EarlyPapers}. One of the first complete models for the origin of UHECR has been based on a particularly powerful subset of RGs \cite{RB93}, and more recently it was shown that also the current results on spectrum and the chemical composition of UHECR can be reproduced with RG sources \cite{EichmannEtAlJCAP2018} (hereafter Paper\,I). The bulk of the contribution is hereby expected to reach Earth nearly isotropically and thus can be described by a continuous source function \cite{Eichmann2019}.

Yet the most challenging test of the hypothesis that RGs are the dominant sources of UHECRs is the comparison with the weak but increasingly significant signatures of anisotropy in the arrival direction distribution of UHECR events. The most important structure to consider here is the dipole detected to $5\sigma$ significance by the Pierre Auger Observatory (hereafter ``Auger'') \cite{Auger_dipole}, but also the less significant indications for intermediate scale anisotropy, seen by Auger and also the Telescope Array experiment (hereafter ``TA'') \cite{int_anis}, have to be taken into account. RGs are particularly suitable to generate such anisotropy as they vary in their power by many orders of magnitude, but to compare this with data a \textit{complete} list of the ``UHECR-brightest'' sources is a precondition. Such a selection has to consider the physics connecting radio flux to UHECR production, as it has been described in Paper\,I, and provide adjustable parameters for all theoretical assumptions made.  We present here a first version of such a ``parametrized UHECR source catalog'' based on radio galaxies and discuss some of its immediate properties. 

\vspace{-1ex}

\section{Source selection}

\noindent For testing models of UHECR origin against observed anisotropies a reliable selection of the \textit{brightest UHECR sources} within the considered model-class is vital. Unfortunately, applying astronomical selection criteria is quite bias-prone, as we can argue for the popular choice of gamma-ray brightness: although there is no doubt that cosmic rays have the potential to produce gamma-rays in interactions, gamma-ray flux (a) depends on the additional presence of a sufficiently dense target population that is not in a simple relation with the cosmic ray density; (b) can also be produced by non-hadronic processes like inverse Compton scattering; and (c) is observed in the GeV-TeV regime, while the cosmic rays we consider here are above EeV. Thus, without a detailed theoretical description how to link the two signals together, any relation between them remains obscure.  
A more robust selection criterion is radio flux, because radio luminosity is known to be in a simple relation to the non-thermal power of an object, which in turn is a plausible scaling quantity for the power in cosmic rays (see also Rachen, these proceedings). Moreover, by relating it to another important non-thermal ingredient of astrophysical sources, magnetic fields, it sets a limit to the highest energy attainable in electromagnetic acceleration (see Paper\,I and references therein). Here we will adopt this relation in a very simple and pragmatic way as the ``scaling paradigm for radio galaxies as UHECR sources'', focusing on the question: Assume RGs are the sources of UHECRs, which objects contribute most?

\subsection{Selection criteria}

\noindent The scaling paradigm for radio galaxies as cosmic ray sources connects the power emitted in cosmic rays at all energies, $L_{\rm cr}$, and the nominal maximum energy $\hat E_{\rm cr}$ up to which 
%protons 
UHECR with charge number $Z$ 
can be accelerated, to observational quantities as 
\begin{equation}
    \begin{array}{lclcl}
         %L_p &\propto& L_{\rm jet} &\propto& (P_1\,d^2)^a \\
         %\hat E_p &\propto& L_{\rm jet}^{1/2} &\propto& g_{\rm s}\,(P_1\,d^2)^{a/2} 
         L_{\rm cr} &\propto& L_{\rm jet} &\propto& (P_1\,d^2)^{\beta_L} \\
         \hat E_{\rm cr} &\propto& L_{\rm jet}^{1/2} &\propto& Z\sqrt{v\,(P_1\,d^2)^{\beta_L}} 
    \end{array}
    \label{eq:LcrEmax}
\end{equation}
Here, $P_1$ is the radio flux of the steep spectrum component (see Sec.\,\ref{sec:boost}) of the source at 1\,GHz, and $d$ the source distance. The power law index $\beta_L$ stems from the relation of radio to jet power, values are assumed to be in the range $0.5{-}0.85$, potentially dependent on the Fanaroff-Riley class of the source (for references see Eichmann~\cite{Eichmann2019}). The impact of the acceleration physics is expressed by the dependency of $\hat E_{\rm cr}$ on a characteristic velocity scale $v$, which is likely higher in the collimated sub-Mpc scale jets of FR-II galaxies ($v\sim 0.3c$) than in FR-I galaxies ($v\sim0.1c$). 

To find out how much a source contributes to the ultra-high energy end of the cosmic ray spectrum, the spectral index $s$ of the cosmic ray spectrum ($dN/dE \propto E^{-s}$) enters as a key parameter. As the cosmic ray spectrum of a UHECR source extends over 10 orders of magnitude, even small variations of this index have huge effects on the energetics. As shown in Paper I, values for $s$ slightly smaller than 2 are preferred to produce the observed UHECR flux with RGs. However, as we just compare individual sources and decide to assign the same value for $s$ to all, we choose $s=2$ for simplicity. Considering then the contribution to cosmic rays above $1\,$EeV observed at Earth for the ``optical case'', i.e., disregarding all propagation effects and assume that their flux is simply $\propto L_{\rm cr} d^{-2}$, we can define a source selection criterion
\begin{equation}
    \label{eqn:selection}
  \left(\frac{P_1/{\rm Jy}}{d/{\rm Mpc}}\right)^{\!\!\frac23}\, \ln\left(\frac{\tilde v\,P_1\,d^2}{14.9\,{\rm Jy\,Mpc^2}}\right) > X\;,
\end{equation}
where the choice of the positive dimensionless number $X$ determines the depth of the catalog. Here, we suppose that the predominant part of UHECRs is composed of protons, i.e.\ $Z=1$, and define $\tilde v \equiv v/0.1c$ as the case for typical FR-I galaxies. We have chosen the index $\beta_L=\frac23$, a value which is supported by both the normalization of jet power on accretion disk properties \cite{RB93} and on kinetic power of the lobes as summarized by Eichmann \cite{Eichmann2019}. The logarithmic term is defined such that the nearest radio galaxy, Centaurus A, has a rigidity cutoff at $\hat E_{\rm cr}/e Z = 7.5\,$EV, and hence can produce the highest observed energies with heavy nuclei ($Z\gtrsim 10$, see Paper I). Note that this term is positive for sources which are powerful enough to accelerate protons up to at least 1\,EeV, given $\tilde v = 1$. 

\subsection{Blazar-like sources and relativistic boosting}
\label{sec:boost}

\noindent Our general scaling relation is based on total jet power, which is released into cosmic rays mostly in the lobes where velocities can be assumed at most weakly relativistic. It is therefore reasonable to assume that RGs emit their cosmic rays nearly isotropic. Nevertheless, observations of blazars, i.e., RGs with their jets aligned to the line of sight according to the common unification paradigm \cite{UrryPadovani}, strongly suggest that radio galaxy jets start up with relativistic velocities. The idea that hadronic processes are active in this core region of the jet has been greatly substantiated by the observation of a cosmic neutrinos likely associated with blazars \cite{Neutrinos}, so we may assume that also cosmic rays emerge from there. As the connection from PeV neutrinos to EeV cosmic rays is still unclear, we choose again a simple pragmatic approach: We assume that, additionally to the isotropic emission of a cosmic ray source, a fraction $f\ll 1$ of this power is emitted in UHECR from the compact jet, with the same spectrum and composition as that of the jet, and then boosted into a narrow cone of opening angle $\Delta\Theta \approx \delta^{-2}$ with the relativistic Doppler factor $\delta=[\Gamma\,(1-\cos\Theta)]^{-1}$, for a typical bulk Lorentz factor $\Gamma\sim 10$. For the borderline case where boosted and unboosted emission are comparable as it is the case for most sources in our sample, the jet must be inclined towards the line of sight by an angle $\Theta \approx 1/\Gamma$, thus $\delta\approx\Gamma$. Assuming ``optical propagation'' again, the flux of cosmic rays is then enhanced by a boosting\footnote{We note that our normalization of UHECR power and maximum energy is done in the lab-frame, thus the boosting considered here is solely an effect of directed emission and not a Lorentz boost from the comoving frame of the jet.} factor $b = 1 + 2\pi f\,\delta^2$, 
% \begin{equation}
%     b = 1+\frac f2\,\frac{4\pi}{\Delta\Theta}=1 + 2\pi f\,\delta^2\,,
%     \label{eq:boosting}
% \end{equation}
where we account for the presence of two jets whereof only one is boosted towards Earth. As a canonical choice, we use $f=0.1$ and $\delta=10$, thus $b \simeq 64$, but we also consider weaker boosting $b=10$ as a conservative choice for selection purposes.

What remains is how sources with ``blazar-like'' properties are selected. A characteristic feature of blazars is a flat ($\alpha\approx 0$) cm-mm spectrum $S_\nu \propto \nu^{-\alpha}$, usually interpreted as a signature of a compact, partially self-absorbed relativistic jet, while for longer wavelengths the steep lobe component (typically $\alpha\approx 0.7$) dominates. Other features revealing blazar-like properties are a compact radio image, significant gamma-ray flux and/or strong variability. For sources showing such features, it is important that the radio flux of the lobes is determined from extending the power law of the low-frequency component to 1 GHz to be used in (\ref{eq:LcrEmax}) for normalization. The boosting is then considered by increasing the resulting $L_{\rm cr}$ by the factor $b$.  

\subsection{The demo-catalog and the strong UHECR source sample}

\noindent As the selection criterion (\ref{eqn:selection}) is difficult to use in standard catalog searches, we first pre-selected radio sources from their listed flux at or around $1\,$GHz. As a base catalog, we used the local radio source catalog by van Velzen et\,al.\,\cite{vV12} (hereafter ``vV12''), as we did in Paper\,I. As noted there already, this catalog suffers from severe deficits in particular in the regime of extremely powerful RGs, because it is based on the 2MRS catalog which selects on infrared flux, and powerful RGs tend to be comparatively dim in infrared. We therefore performed additionally a search for objects with radio flux ${>}3\,$Jy (and \textit{no} other criteria applied) in NED\footnote{NASA/IPAC extragalactic database, https://ned.ipac.caltech.edu/}, and then drew our sample from the joint list by selecting confirmed RGs with a  distance ${<}\,300\,$Mpc, which was found to be a reasonable maximum distance for UHECR propagation above 10\,EeV \cite{Stanev:2000}. This way we obtained a list of 42 sources\footnote{This choice is based on the well-known fact that 42 is the answer to everything, so we cannot do wrong using it.}, which all fulfill (\ref{eqn:selection}) for $X=1.3$ (for boosted sources this applies for $b\ge 10$).      

\begin{table}[t]
    \centering
    \begin{footnotesize}
    \renewcommand{\thefootnote}{\fnsymbol{footnote}}
    \newcommand{\fm}[1]{\footnotemark #1}
    \rmxs
    \renewcommand{\arraystretch}{0.8}
    \begin{tabular}{l|rr|r|r|l||l|rr|r|r|l}
Radio name    	& RA  	      	&  DEC 	   & $\scriptstyle P_1$/Jy	& $\scriptstyle d$/Mpc     	& type &
Radio name    	& RA  	      	&  DEC 	   & $\scriptstyle P_1$/Jy	& $\scriptstyle d$/Mpc     	& type\\\hline	
3C\,31          & 16.85         &  32.41        & 7             & 74            & FR-I & % checked, X=1.6
3C\,33          & 17.22         &  13.34        & 16            & 260           & FR-II\\ % checked, X=1.7
3C\,40          & 21.50         &  -1.34        & 8             & 77            & FR-I & % checked, X=1.8
PKS\,0131-36    & 23.49         & -36.49        & 9             & 130           & FR-I/II\\ % checked, X=1.6
3C\,66B         & 35.79         &  42.99        & 12            & 92            & FR-I & % checked, X=2.3
3C\,66          & 35.61         &  43.01        & 12            & 291           & FR-I\\ % checked, X=1.3
PKS\,0238-084   & 40.27         &  -8.26        & 0.5           & 21            & BLU & % checked, X=0.22 (14)
3C\,78          & 47.11         &   4.11        & 9             & 123           & FR-I\\ % checked, X=1.6
3C\,83.1        & 49.57         &  41.85        & 10            & 109           & FR-I & % checked, X=1.8
Perseus\,A\fm[1]\tiny{$^b$}   & 49.95 &  41.51        & 7             & 76            & BLU\\ % checked, X=1.6 (100)
Fornax\,A\fm[1]               & 50.67 & -37.21        & 163           & 25            & FR-I & %checked, X=31
3C\,98          & 59.73         &  10.43        & 13            & 132           & FR-II\\ % checked, X=2.1
3C\,111\fm[1]\tiny{$^b$}      & 64.59 &  38.03        & 19            & 212           & FR-II/SSRQ & % checked, X=2.2 (140)
3C\,120\fm[1]\tiny{$^b$}             & 68.30 &   5.35        & 3             & 143           & BLU\\ % checked, X=0.6 (40)
3C\,129         & 72.29         &  45.01        & 10            & 91            & FR-I & % checked, X=2.0
3C\,134         & 76.18         &  38.10        & 13            & ---\tiny{\fm[2]}     & FR-II\\ % checked, X=1.4-3.8
Pictor\,A\fm[1]               & 79.96 & -45.78        & 80            & 155           & FR-II & % checked, X=7.6
PKS\,0521-36\fm[1]\tiny{$^b$} & 80.74 & -36.46        & 21            & 249           & SSRQ\\ % checked, X=2.2 (140)
% 4C +56.16     & 117.15        &  55.82        & 6             & 159           & FR-II\\ % checked, X=1.0*
Hydra\,A\fm[1]  & 139.52        & -12.10        & 62            & 247           & FR-I & % checked, X=5.0
3C\,264         & 176.27        &  19.61        & 8             & 104            & FR-I\\ % checked, X=1.5
3C\,270\fm[1]   & 184.85        &   5.82        & 23            & 32            & FR-I & % checked, X=5.9
3C\,272.1       & 186.27        &  12.89        & 8             & 15            & FR-I\\ % checked, X=3.1
Virgo\,A\fm[1]  & 187.71        &  12.39        & 283           & 15            & FR-I & % checked, X=60
PKS\,1245-41    & 192.21        & -41.31        & 5             & 43            & FR-I\\ % checked, X=1.5 
3C\,278         & 193.65        & -12.56        & 10            & 75            & FR-I & % checked, X=2.1
Centaurus\,A\fm[1]    & 201.37  & -43.02        & 300           & 4             & FR-I\\ % checked, X=100
PKS\,1333-33    & 204.16        & -33.97        & 11            & 45            & FR-I/II & % checked, X=2.9
Centaurus\,B\fm[1]    & 206.70  & -60.41        & 140           & 54            & FR-I\\ % checked, X=19
3C\,310         & 226.24        &  26.02        & 11            & 246           & FR-I & % checked, X=1.4* 
3C\,317         & 229.19        &   7.02        & 8             & 162           & FR-I\\ % checked, X=1.3* 
PKS\,1610-60\fm[1]    & 243.77  & -60.91        & 70            & 78            & FR-I & % checked, X=9.5
4C\,+39.49      & 253.47        &  39.76        & 0.5           & 155           & BLO\\ % checked, X=0.15 (9.4)
% TXS\,1710-249   & 258.31        & -25.04        & 6             & 135           & FR-I & % checked, X=1.1*
3C\,353\fm[1]   & 260.11        &  -0.97        & 75            & 143           & FR-II & % checked, X=7.5
3C\,371         & 271.71        &  69.82        & 1             & 228           & BLU \\ % checked, X=0.21 (14)
PKS\,1814-63\fm[1]\tiny{$^b$}    & 274.90  & -63.76    & 18            & 290           & BLU & % X=1.8 (115) 
3C\,386         & 279.61        &  17.20        & 9             & 81            & FR-I\\ % checked, X=1.9 
3C\,390.3       & 280.54        &  79.77        & 15            & 250           & FR-II & % checked, X=1.7
Cygnus\,A\fm[1] & 299.87        &  40.73        & 2225          & 250           & FR-II\\ %checked, X=69
% PKS\,2058-28    & 315.41        & -28.41        & 8             & 177           & FR-I/II\\ % checked, X=1.2*
PKS\,2104-25    & 316.86        & -25.42        & 17            & 173           & FR-I & % checked, X=2.2
3C\,430         & 319.58        &  60.80        & 10            & 246           & FR-II\\ % checked, X=1.3 
PKS\,2153-69\fm[1]    & 329.27  & -69.69        & 35            & 128           & FR-I/II & % checked, X=4.5
3C\,465         & 354.62        &  27.03        & 11            & 132           & FR-I\\\hline % checked, X=1.8
\multicolumn{12}{l}{\footnotemark[1]Object included in the strong UHECR source sample; for sources marked with a \tiny{$b$} this applies only if boosting is applied with \tiny{$b\ge 10$}.\normalsize\strut }\\
\multicolumn{12}{l}{\footnotemark[2]Object with unknown redshift. It was included as its observed size and radio flux suggests a distance lower than 300\,Mpc.}
    \end{tabular}
    \end{footnotesize}
    \caption{The 42 sources of the demo catalog, with objects included in the strong UHECR source sample marked. Sources in this sub-sample can significantly influence UHECR anisotropy. Sources are generally named after radio-catalogs and radio naming conventions, in the order \textit{Constellation (Con)} A/B, 3C, PKS, 4C, although some are better known under other names, e.g., 4C\,+39.49 $\equiv$ Mrk\,501. The ``type'' column lists the Fanaroff-Riley type including borderline cases, and for beamed sources ``BLO'' for BL Lac objects, ``BLU'' for FR-I based blazars with a significant steep spectrum component, and SSRQ for Steep-Spectrum Radio Quasars, which is the corresponding FR-II based source class \cite{UrryPadovani}.}
    \label{tab:smaple}
\end{table}

In an a posteriori analysis, considering UHECR transport as described in the next sections, it turned out that the sample naturally falls into two parts: sources with $X < 3$ which have only a minor impact on anisotropy, and sources with $X > 4$. The latter we use as a criterion to define a sub-sample of ``strong UHECR sources'' (where again for boosted sources the selection criterion is applied for $b\ge 10$), and it can be considered complete in the sense that (a) there exists no radio galaxy within $300\,$Mpc with a comparable impact on UHECR anisotropy, (b) omission of sources not contained in it will hardly affect the results. We particularly note that 6 objects out of this list of 16 --- Pic\,A, 3C\,111, Cen\,B, PKS\,0521-36, PKS\,1610-60 and PKS\,1814-63 --- are \textit{not} contained in vV12. Table\,1 shows the full demo-catalog with the strong source sub-sample marked.

\section{Considering UHECR transport}

\noindent The selection criteria applied so far treated cosmic rays as protons that propagate like radio photons -- we called this the ``optical scenario''. Of course, this view cannot hold for three reasons: (a) the presence of heavy nuclei in the UHECR spectrum, as implied by Auger results \cite{Auger_Comp}, (b) energy losses of UHECRs and photodisintegration of nuclei in interactions with ambient extragalactic photon backgrounds, and (c) deflection of cosmic rays in extragalactic magnetic fields (EGMF). All these effects can be ideally treated in simulations with the cosmic ray propagation code CRPropa \cite{CRPropa3}, where we use a set of possible EGMF models provided by Hackstein et al.\ \cite{Hackstein2018}, hereafter referred to as H+18a, H+18aR, H+18p, and H+18p2R using the same naming convention as Eichmann (these proceedings, see also for details). 

\subsection{Simulation setup and weighting}

\noindent In order to obtain sufficient statistics from distant sources for all H+18 models within a reasonable CPU time we use the so-called inverted simulation setup introduced by Eichmann \cite{Eichmann2019}. As the H+18 models are limited to a volume that does not include sources at distances $\gtrsim 125\,$Mpc, we reflect the magnetic field structure at its boundaries, based on the assumption of a homogeneous Universe on large scales. 
%This approach can not account for the proper spatial distribution of the EGMF, but the impact of the chosen EGMF model is significantly higher than the proper spatial positions of the sources and the observer, in particular in the case of a source distance that is significantly larger than the particle's Larmor radius. 
%Thus, we use this simulation method and place an isotropically emitting source at the centre of multiple observer spheres, whose radii are determined by the proper source distance as given in table \ref{tab:smaple}. 
An isotropically emitting source is placed at the center of ten concentric observer spheres, with radii chosen such that they represent all source distances within $\approx 10\%$.
Each observer sphere records the CR properties each time it passes through its surface in any direction, until the particle is removed from the simulation when either the trajectory length exceeds $5000\,$Mpc or its energy dropped to $<1\,$EeV. To obtain about equal statistics within the simulated energy range as well as the chemical composition, the source ejects a solar composition of particles enhanced by $Z^{2.5}$ with an energy spectrum between $1\,$EeV and $300\,Z\,$EeV with a spectral index $s=1$. To exclude the impact of a particular source position in the EGMF structure, we run 50 different positional setups of $10,000$ particles for each of the H+18 EGMF models. 

From these data, the cosmic ray contribution of individual sources is obtained via (\ref{eq:LcrEmax}), where we also re-weight to an $s=2$ spectrum with $Z^2$ enhancement (see Paper I). In the case of blazar-like sources, particles that are emitted within an opening angle $\Delta\Theta=0.1$ with respect to the line of sight obtain an additional boosting weight $b=64$. To consider a limited source age $T_s$, we select only particles with trajectory lengths $d$ on a sphere with radius $r$ which satisfy $d-r < c\,T_s$. We account for the angular dependence of effective area $dA$ in our setup by applying a weight $1/|\cos\theta|$.

%can be used to estimate the mean deflection .

%\subsection{Weighting sources using simulation results}
%
%not needed? already in previous section!?!

\section{Sky maps and comparison to anisotropy signals}

\noindent The results presented here demonstrate some immediate properties of our source selection, useful to determine in practical applications whether more detailed and expensive simulations make sense for a specific parameter set, but should not be taken as definite predictions of the ``radio galaxy model'' suitable to confirm or refute it. The maps are in galactic coordinates, and when we refer to individual sources in the text we add $(l,b)$ in degrees to allow their easy identification in the maps.

{\setlength{\textfloatsep}{0pt}

\subsection{UHECR emission power and flux contribution}

\noindent The upper left panel of Fig.\,1 shows the source sample in symbols with their area proportional to their cosmic ray power \textit{ejected} above 4\,EeV in our direction. This figure shows that besides the ``usual suspects'' Cen\,A $(310,+19)$, Vir\,A $(284,-74)$, For\,A $(240,-57)$ and Cyg\,A $(76,+6)$, for the case of canonical boosting also Per\,A $(151,-13)$, 3C\,111 $(162,-9)$, PKS\,0521-36 $(241,-33)$ and PKS\,1814-63 $(331,-21)$ are expected to contribute to the UHECR flux on a similar level. In its upper right panel, the same sky plot is shown, but this time with symbol sizes weighted by \textit{observed} particle flux. Here and in the following we consider UHECR transport for H+18aR fields. The contribution of distant sources is at this low energy diminished mostly because particles are delayed in their arrival time by more than $100\,$Myr. We have applied this upper bound as RGs are in fact transient objects on these time scales, but in special cases it might be shorter: For Cyg\,A we used a maximum delay of $50\,$Myr to account for the recent results of Pyrzas et al.\ \cite{PSB2015}. Switching to the weaker H+18a fields yields a similar picture, while essentially no cosmic rays are received from \textit{any} distant source in the H+18p/p2R scenarios.     

\begin{figure}[tb]
    \centering
    \begin{footnotesize}
    % \vspace{0.18\textheight}
    \includegraphics[width=0.49\textwidth, %height=0.3\textheight, 
    viewport= 206 670 1126 1142, clip]{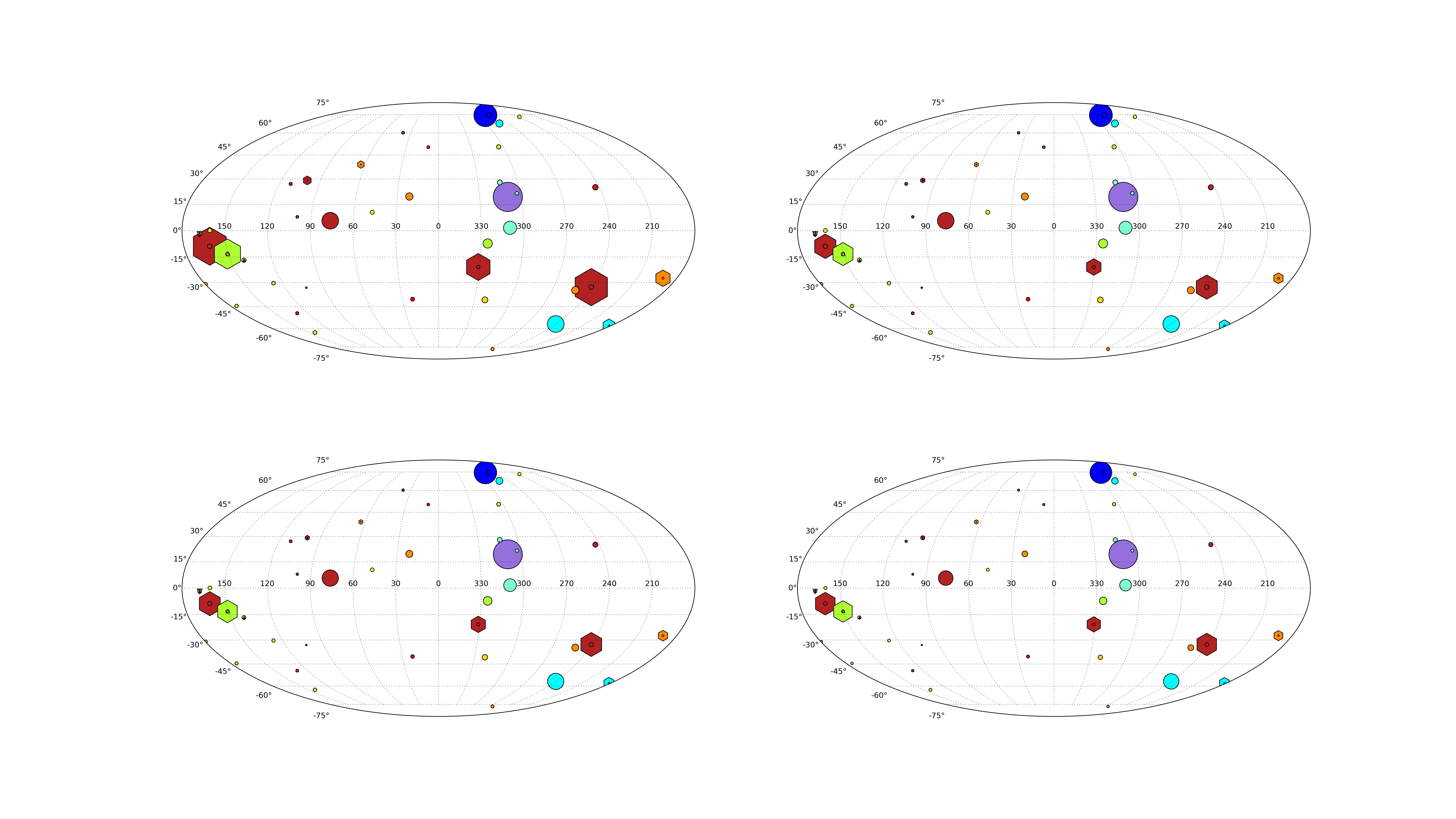}%
    \includegraphics[width=0.49\textwidth, %height=0.3\textheight,
    viewport= 1192 670 2102 1142, clip]{RGmap_H18aR_4EeV.png}
    % \vspace{0.18\textheight}
    \includegraphics[width=0.49\textwidth, %height=0.3\textheight,
    viewport= 206 95 1126 569, clip=True]{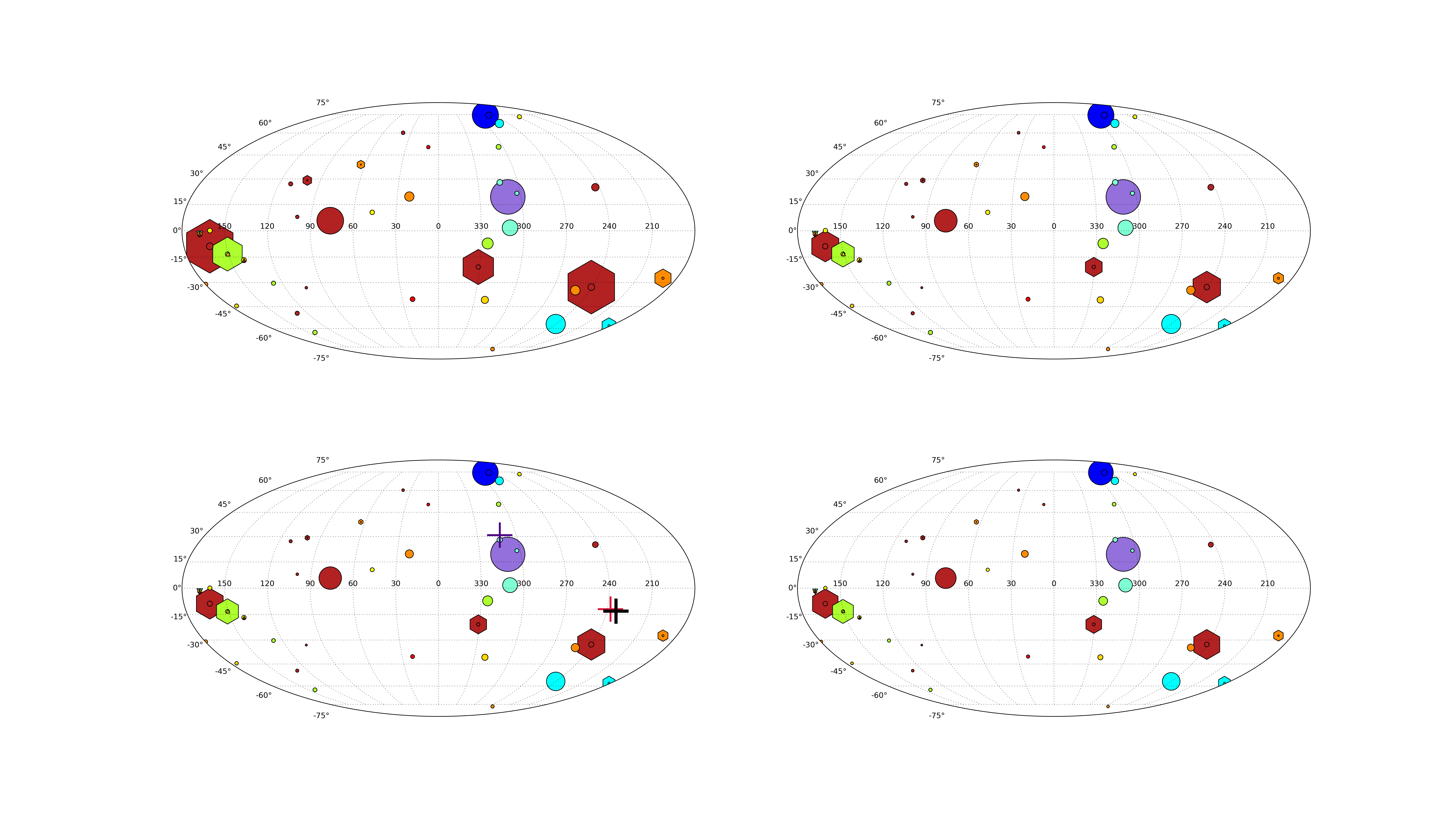}%
    \includegraphics[width=0.49\textwidth, %height=0.3\textheight,
    viewport= 206 95 1126 569, clip=True]{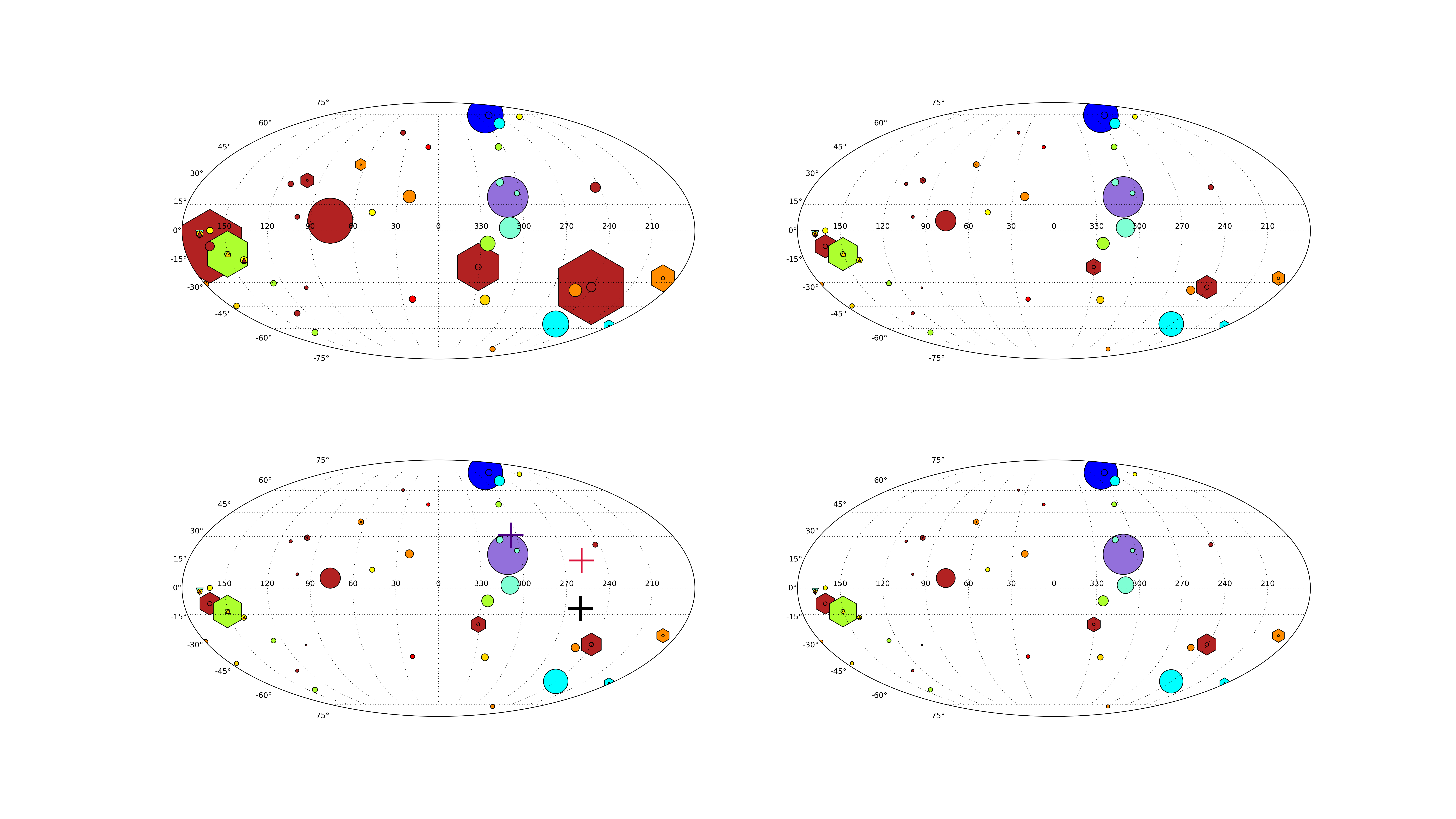}
    \caption{Map of radio galaxies as UHECR sources, weighted by emitted (upper left) and observed (upper right) cosmic rays above 4 EeV. Unboosted emission of each source is shown by a circle, the relative contribution scales with the circle area -- in some cases, where two source are too close in their lines of sight, triangles are used to distinguish them. For blazar-like sources with the potential of boosted emission, hexagons indicate their contribution for a canonical boosting factor given in text. Colors indicate the source distance, from purple for the closest source (Cen\,A) to deep-red for the most distant ones. The lower panel shows the dipole contribution above 8 EeV (left) and 32 EeV (right). The total dipole is calculated from adding the dipole vectors of all sources, with (red) or without (purple) consideration of boosting for blazar-like sources. Black are the positions of the Auger dipoles for these energies \cite{Auger_dipole}. By the way, the tiny hexagon at (64,+39) is Mrk\,501, the strongest contributor among TeV blazars.}
    \end{footnotesize}
    \label{fig:uhecr_emitted_observed}
\end{figure}

\subsection{Dipole contribution and hot spots in directed flux}

\noindent Turning to anisotropy signatures in the UHECR flux, the lower panel of Fig.~1 shows the dipole contribution of our source sample above 8 and 32\,EeV, where the symbol sizes scale with energy according to the source spectrum multiplied with $E^{2.7}$. We see that for ${>}\,8\,$EeV the direction of its total dipole is in excellent agreement with Auger results, mostly caused by the strong contributions of 3C\,111, Per\,A and PKS\,0521-36, and still in reasonable agreement for ${>}\,32\,$EeV. We note, however, that this agreement stands and falls with the validity of the ``canonical boosting'' argument -- if ejected cosmic rays from the core jet are fanned out by magnetic fields in the lobes, boosting will be much weaker and ultimately all anisotropy will be drawn towards the direction of Cen\,A -- the purple crosses mark the positions of the dipoles is boosting is disregarded altogether.  
% The dipole strength of our source sample is generally too high, but we note that this would be reduced by an expected strong and almost isotropic contribution from background radio galaxies, as discussed by Eichmann (these proceedings). 
% \begin{figure}[tb]
%     \centering
%     \begin{footnotesize}
    
%     \caption{
%     Symbol sizes and colors are as in fig.~\ref{fig:uhecr_emitted_observed}.}
%     \end{footnotesize}
%     \label{fig:dipole}
% \end{figure}
}

{\setlength{\textfloatsep}{0pt}

Finally, Fig.\,2 shows the situation for cosmic rays observed above 57\,EeV, where we compare to the ``hot spots'' in directional flux reported by both leading experiments. It should be mentioned that our results do not include UHECR deflection in the Galactic magnetic field (GMF), which could reach up to $30^\circ$ or more for intermediate mass nuclei \cite{FS2019}.\footnote{%
The reason why we did not ``simply'' apply GMF deflection maps contained in CRPropa is that creating sky maps of arrival directions is beyond the scope of this paper, and refer to the initial paragraph of this section.} 
With this in mind, we emphasize that the strongest and most significant excess reported by Auger is almost identical to the position of the nearest strong radio galaxy, Cen\,A, while the TA hot spot and the Auger excess near the Galactic south pole are at about $30\degree$ distance of the strong RGs Vir\,A and For\,A, respectively. The contribution of distant sources is at these high energies strongly diminished by the GZK effect. 
\begin{SCfigure}
    \begin{footnotesize}
    \includegraphics[width=0.49\textwidth, %height=0.3\textheight,
    viewport= 1192 125 2102 569, clip]{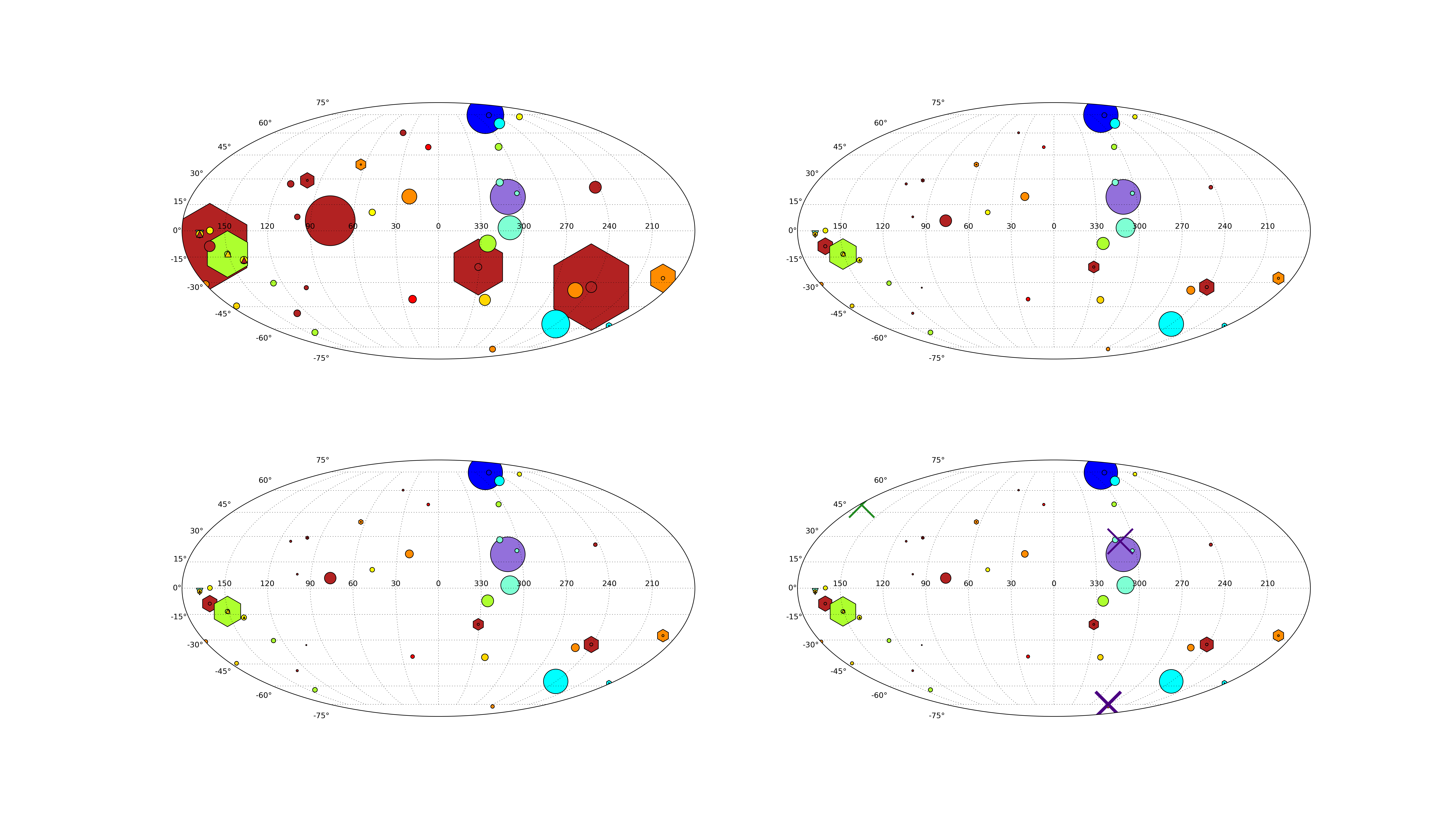}
    \caption{Map of radio galaxies as UHECR sources, weighted by their contribution to intermediate scale (${\sim}\,10\degree$) anisotropy. Crosses show the approximate position of flux excesses reported by Auger (blue) and TA (green) \cite{int_anis}. Symbol sizes and colors are as in Fig.~1.
    
    \strut}
    \end{footnotesize}
\label{fig:hotspots}
\end{SCfigure}

\section{Conclusions and Outlook}

\noindent We presented a first demo version of a UHECR source catalog based on radio galaxies, including a ``strong UHECR source sample'' of 16 radio galaxies which can be regarded as the currently most reliable and complete selection of sources which could contribute to UHECR flux anisotropy on a level testable with current statistics. Simple and efficient tests of the sample properties for canonical parameters demonstrate that radio galaxies have the potential to explain all currently detected or claimed UHECR anisotropies. We will soon provide an extended and completed version of this catalog in electronic form, most likely as an extension of the {\sc Source}-Class in CRPropa, to be used in high quality simulations which can be compared with actual and future data.
}

% \begin{footnotesize}
% \bibliographystyle{plain}
% \bibliography{references}
% \end{footnotesize}

\end{document}